\documentclass[11pt,twoside]{article}
\usepackage{asp2010}

\resetcounters

\begin{document}

\title{The Bipolar X-Ray Jet of the Classical T Tauri Star DG Tau}
\author{M. G\"udel$^1$, M. Audard$^2$, F. Bacciotti$^3$, J.~S. Bary$^4$, K.~R. Briggs$^5$, 
        S. Cabrit$^6$, A. Carmona$^2$, C. Codella$^3$, C. Dougados$^7$, J. Eisl\"offel$^8$, 
	F. Gueth$^9$, H.~M. G\"unther$^{10}$, G. Herczeg$^{11}$, P. Kundurthy$^{12}$, 
	S.~P. Matt$^{13}$, R.~L. Mutel$^{14}$, T. Ray$^{15}$, J.~H.~M.~M. Schmitt$^{16}$, 
	P.~C. Schneider$^{16}$, S.~L. Skinner$^{17}$, R. van Boekel$^{18}$
\affil{$^1$University of Vienna, Dept. of Astronomy, T\"urkenschanzstr. 17, A-1180 Vienna, Austria}
\affil{$^2$ISDC Data Centre for Astrophysics \& Observatoire de Gen\`eve, University 
           of Geneva, Ch. d'Ecogia 16, CH-1290 Versoix, Switzerland}
\affil{$^3$INAF, Osservatorio Astrofisico di Arcetri, Largo E. Fermi 5, 50125 Firenze, It}
\affil{$^4$Colgate University, Dept. of Physics and Astronomy, 13 Oak Drive, Hamilton, NY 13346, USA}
\affil{$^5$ETH Zurich, Institute of Astronomy, Wolfgang-Pauli-Str. 27, 8093 Zurich, Switzerland}
\affil{$^6$L'Observatoire de Paris, 61, avenue de l'Observatoire, 75014 Paris, France}
\affil{$^7$Laboratoire d'Astrophysique de Grenoble, UMR 5571, BP 53, 38041 Grenoble Cedex 09, France}
\affil{$^8$Th\"uringer Landessternwarte, Sternwarte 5, D-07778 Tautenburg, Germany}
\affil{$^9$Institut de RadioAstronomie Millim\'etrique, 300 rue de la Piscine, Domaine Universitaire, 
         38406 Saint Martin d'H\`eres, France}
\affil{$^{10}$Harvard-Smithsonian Center for Astrophysics, 60 Garden Street, Cambridge, MA 02138, USA}
\affil{$^{11}$Max Planck Institut für Extraterrestrische Physik, Giessenbachstrasse 1, 85748 Garching, Germany}
\affil{$^{12}$University of Washington, Dept. of Astronomy, Seattle, WA 98195-1580, USA}
\affil{$^{13}$Laboratoire AIM Paris-Saclay, CEA/Irfu Universit\'e Paris-Diderot CNRS/INSU, 91191 Gif-sur-Yvette, France}
\affil{$^{14}$Dept. of Physics and Astronomy, University of Iowa, Iowa City, Iowa, USA}
\affil{$^{15}$School of Cosmic Physics, Dublin Institute for Advanced Studies, Dublin 2, Ireland}
\affil{$^{16}$Hamburger Sternwarte, Gojenbergsweg 112, 21029 Hamburg, Germany}
\affil{$^{17}$CASA, 389 UCB, University of Colorado, Boulder, CO 80309-0389, USA}
\affil{$^{18}$Max-Planck-Institute for Astronomy, K\"onigstuhl 17,  69117 Heidelberg, Ge}}

\begin{abstract}
We report on new X-ray observations of the classical T Tauri star DG Tau.
DG Tau drives a collimated bi-polar jet known to be a source of X-ray emission 
perhaps driven by internal shocks. The rather modest extinction permits
study of the jet system to distances very close to the star itself. Our initial 
results presented here show that the spatially resolved X-ray jet has been moving 
and fading during the past six years. In contrast, a stationary, very soft 
source much closer ($\approx 0.15-0.2^{\prime\prime}$) 
to the star but apparently also related to the jet has brightened
during the same period. We report accurate temperatures and absorption column
densities toward this source, which is probably associated with the jet
base or the jet collimation region.
\end{abstract}

\section{Introduction}

DG Tau is a classical T Tauri star (CTTS) with a flat infrared spectrum, indicating the 
presence of substantial circumstellar material. DG Tau ejects a well-studied bipolar jet, 
showing several knots and shocks (e.g., \citealt{eisloeffel98}).
The most prominent knots are presently located about 5$^{\prime\prime}$ and 12$^{\prime\prime}$ 
away from the star, toward the SW. The NE counter-jet is difficult to see owing to absorption/extinction
by the foreground extended disk structure \citep{kitamura96}. We discovered DG Tau's 
jet in X-rays (\citealt{guedel05, guedel07, guedel08}; G05, G07, resp. G08 henceforth) along with a spectral anomaly in the 
central source also ascribed to the jet. This anomaly manifests itself in a superposition of two 
unrelated spectral components subject to different hydrogen absorption column densities, $N_{\rm H}$, in 
the unresolved central point spread function (PSF) that contains the star itself (defining a ``Two-Absorber X-Ray Spectrum'' = TAX,
G07). The hard component is ascribed to the flaring corona/magnetosphere of the star, excessively 
absorbed by dust-depleted accretion streams, while the little absorbed soft component is ascribed to 
X-ray emission from the jet base. We have started a multi-wavelength campaign to study DG Tau in detail, 
centered around a Chandra Large Program (360~ks of ACIS-S time in January 2010, complementing our
90 ks obtained in 2004-06) and involving radio, millimeter, infrared, and optical telescopes. 
Here, we present initial results, focusing on the X-ray phenomenology.

\section{Jet Morphology and Kinematics}
\begin{figure}[t!]
\centerline{\includegraphics[angle=0,width=9.cm]{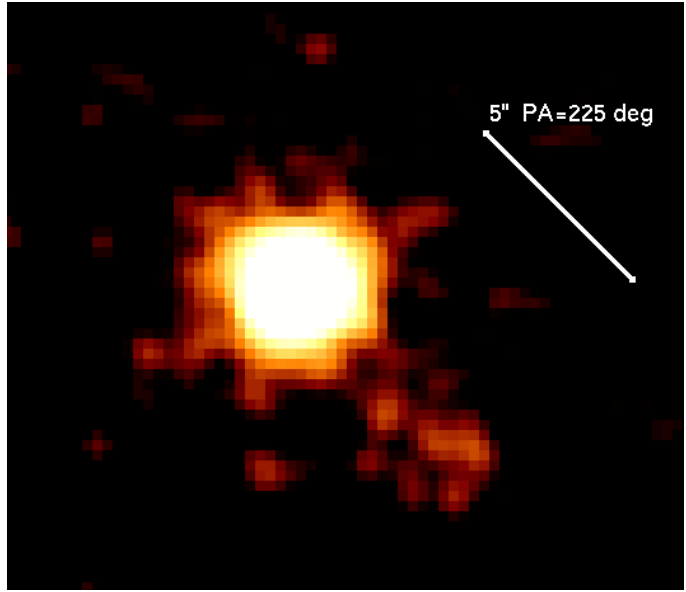}}
\caption{X-ray jet of DG Tau observed in January 2010, in the energy range 0.4--1.5~keV; the original
data have been smoothed using a Gaussian kernel of width 1.2 pixels, where 1 pixel = 0.25$^{\prime\prime}$.}
\label{fig1} 
\end{figure}

The DG Tau X-ray jet is visible at PA=225 deg out to a distance of 5-6$^{\prime\prime}$. 
Figure~\ref{fig1} is based on the January 2010 observations (360 ks). To optimize spatial resolution, 
we reconstructed events files for the detector's ``VFAINT'' mode (allowing for better background treatment), 
removing standard pixel randomization and applying 
the subpixel event repositioning (SER) method by \citet{li03} and \citet{li04}. The image has been smoothed 
using a Gaussian kernel of width 1.2 pixels (1 pix = 0.25$^{\prime\prime}$), and intensity has been 
logarithmically compressed, including only 0.4-1.5~keV counts. The jet to the SW appears to be brightest 
toward its apex. This source coincides with an optical knot (see below). It is therefore likely 
that the gas has been heated by shocks forming in the jet. 
\begin{figure}[b!]
\centerline{
\hbox{
\includegraphics[angle=0,width=6.5cm]{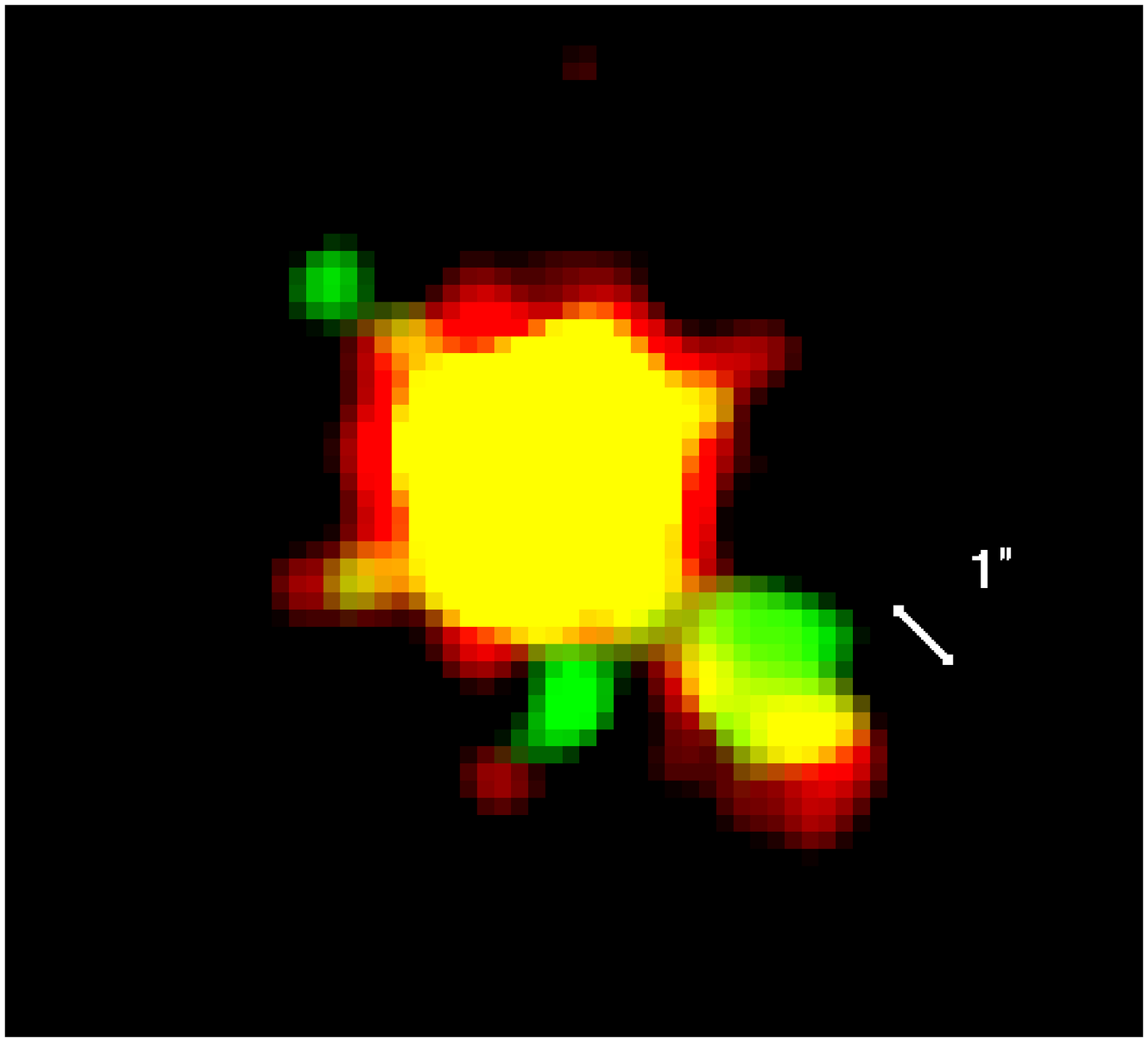}
\includegraphics[angle=0,width=6.1cm]{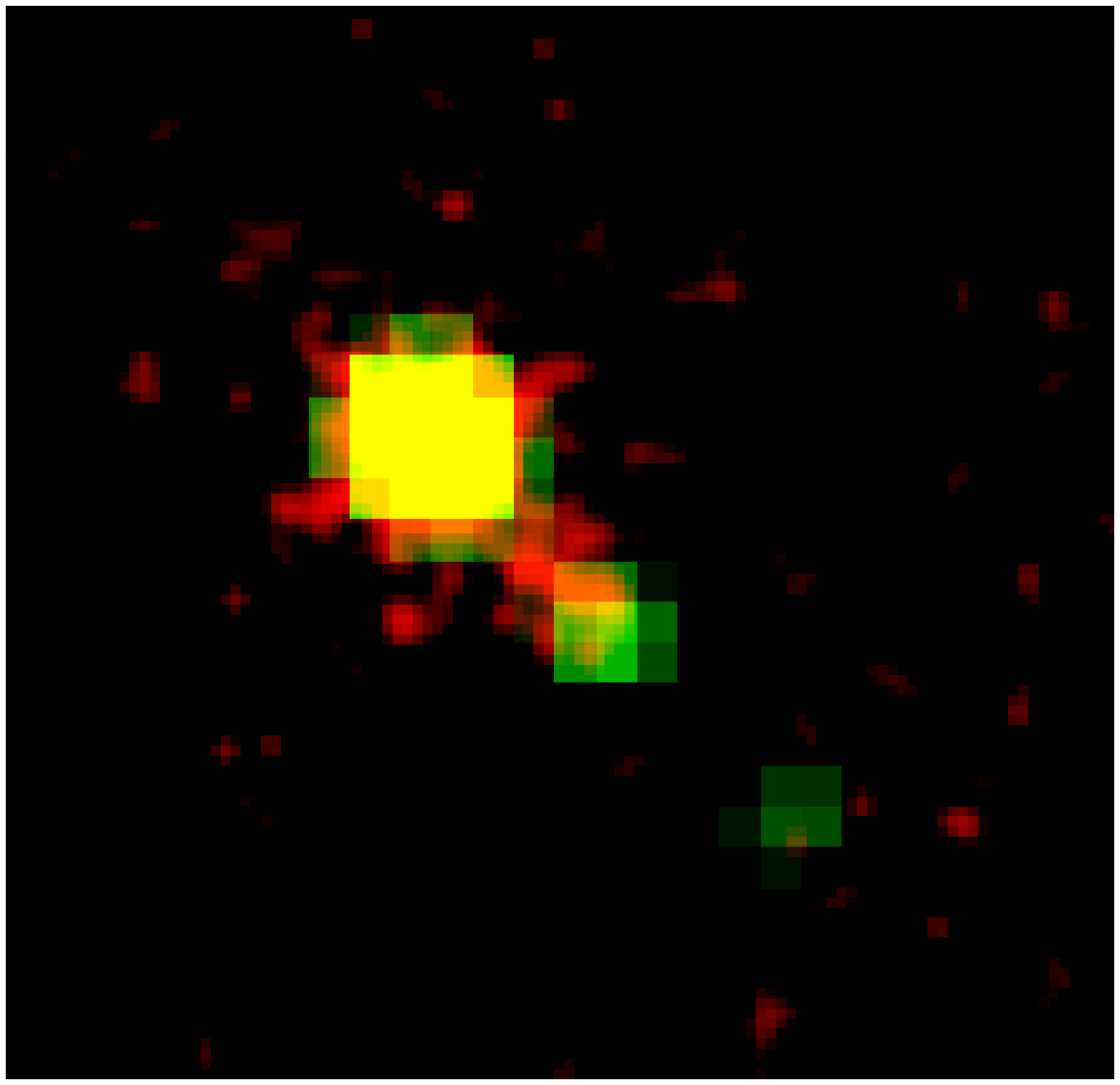}
}
}
\caption{{\it Left} (a): Two-color image, showing superposition of smoothed X-ray images for Winter 
          2005/06 (green) and for January 2010 (red), indicating jet longitudinal motion.
-- {\it Right} (b): Calar Alto Potsdam Multi Aperture Spectrophotometer (PMAS) [S\,{\sc ii}] image 
  (green) superimposed on a smoothed X-ray image (red).
}
\label{fig2} 
\end{figure}

The optical jet expands by 
0.15-0.3$^{\prime\prime}$~yr$^{-1}$ \citep{eisloeffel98, dougados00}. The 2-color image 
in Fig.~\ref{fig2}a shows a superposition of the (smoothed) X-ray data from Winter 2005/06 (green, 60 ks) and January 2010 (red). 
Jet motion is discernible. Using the wavdetect task in CIAO (on unsmoothed data), we find a velocity 
of $\approx 0.2^{\prime\prime}$~yr$^{-1}$ toward PA = 225 deg, coincident with the optical 
velocity. The X-ray shock regions are thus co-moving with the jet and are clearly not standing shocks.

The X-ray jet apparently faded between 2004 and 2010. X-ray count rates in identical areas of the
SW jet were 0.20$\pm$0.08~ct~ks$^{-1}$ (1$\sigma$, 2004), 0.18$\pm$0.06~ct~ks$^{-1}$ (2006), and
0.11$\pm$0.02~ct~ks$^{-1}$ (2010), indicating a marginal trend consistent with cooling estimates in 
G08 for rather high electron densities (e.g., 10$^5$~cm$^{-3}$).

Figure~\ref{fig2}b shows a superposition of X-rays (red) with blueshifted [S\,{\sc ii}] 
(-250 km~s$^{-1}$ to -100~km~s$^{-1}$, green) observed with PMAS (Potsdam Multi Aperture Spectrophotometer) 
attached to the 3.5~m telescope at Calar Alto (mean seeing was 1.8$^{\prime\prime}$, pixel size
$1^{\prime\prime}\times 1^{\prime\prime}$, observing date 16 November 2009). The [S\,{\sc ii}] knot is located 
at the tip of the soft X-ray jet.

\section{The Counter-Jet}

\begin{figure}[t!]
\includegraphics[angle=0,width=13.cm]{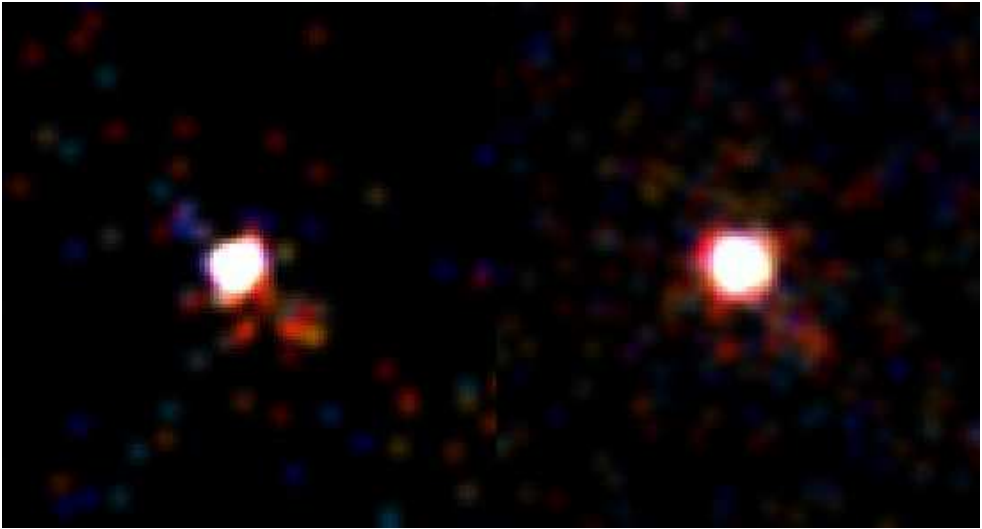}
\caption{Smoothed hardness images of the DG Tau jet in 2004-2006 (left) and 2010 (right). Red is softest, blue hardest
(for photons ranging from 0.5-1.5~keV). Note the absence of a significant counter-jet in 2010.}
\label{fig3} 
\end{figure}

The 2004-06 Chandra data (90~ks) showed a clear X-ray detection of the counter jet 
(Fig.~\ref{fig3}, left; 9 counts, see G08). Although still dominated by soft photons, its spectral appearance is harder 
than the forward jet. This has been ascribed to photoelectric absorption by the foreground extended
disk. The excess $N_{\rm H}$ (compared to the forward jet) was found to be $3\times 10^{21}$~cm$^{-2}$,
compatible with the excess extinction toward the counter jet (G08). The image also shows --
marginally -- that the forward jet may be harder at its apex; this may be a sign of shock heating and 
post-shock cooling.

The counter-jet is hardly seen in the 4x longer 2010 
exposure (Fig.~\ref{fig3}, right) although we had expected to find 36 counts. The cause is unclear; most 
likely the jet has cooled  by expansion  and radiation (which is true for the forward jet as well -- see below).

\section{Closer to the Star: The Inner-Jet}

\begin{figure}
\hbox{
\includegraphics[angle=0,width=13cm]{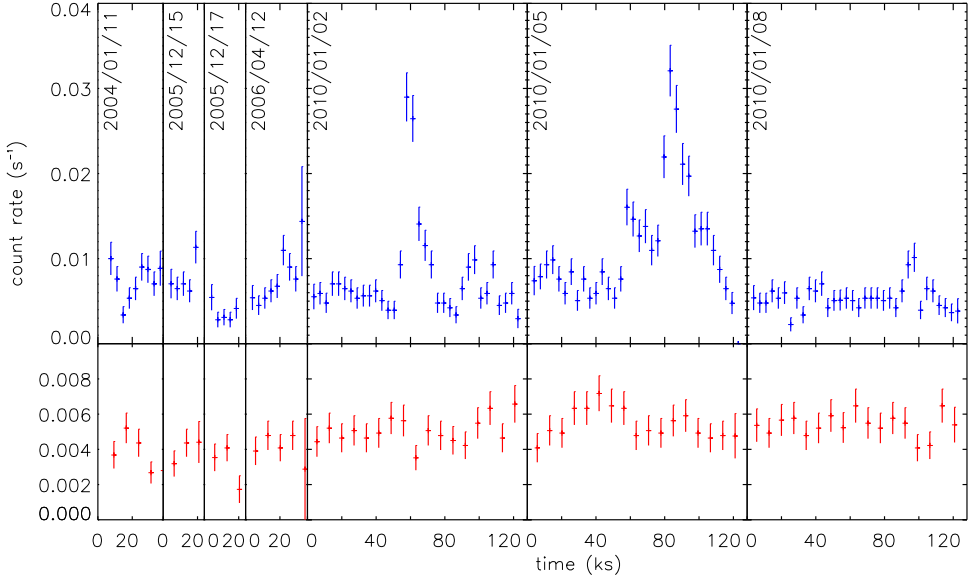}
}
\caption{Hard (1.5--7.3~keV. upper panel) and soft (0.5-1.0~keV, lower panel) light curves extracted from the central PSF of 
         all Chandra DG Tau observations; the individual observations are separated by vertical lines,
	 and observing dates are given in the upper panels. Note the absence of correlated behavior in
	 the two light curves and the nearly constant flux level in the soft emission except for a gradual
	 increase in luminosity over 6 years, from $1.1\times 10^{29}$~erg~s$^{-1}$  to $1.8\times 10^{29}$~erg~s$^{-1}$.}
\label{fig4} 
\end{figure}

\begin{figure}
\centerline{
\includegraphics[angle=-90,width=11.cm]{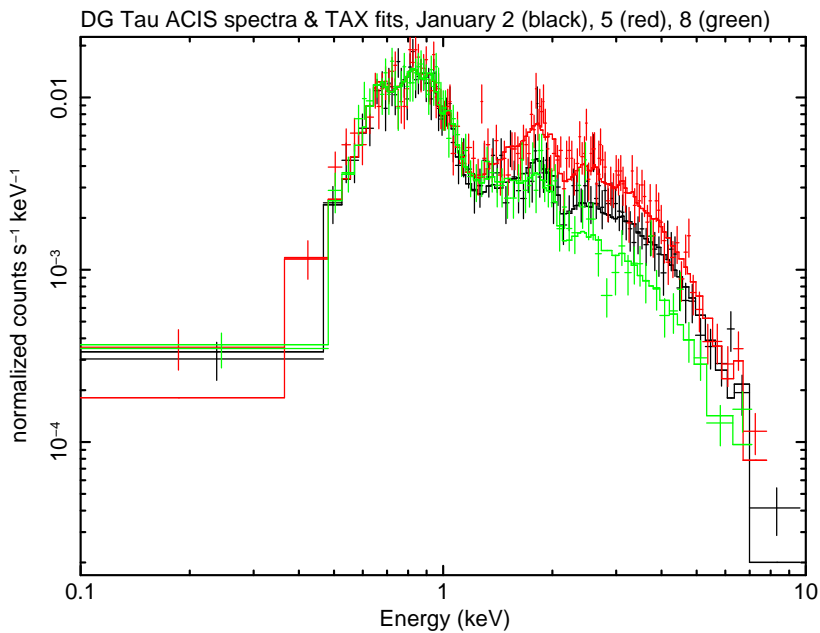}
}
\caption{Chandra ACIS-S spectra of the DG Tau X-ray source in the central PSF. The spectra refer to observations taken
          on January 2 (black), January 5 (red), and January 8 (green), 2010. Note variability above 1.5~keV due
	  to flares (see Fig.~\ref{fig4}), but identical spectra below $\approx$1.2~keV.}
\label{fig5} 
\end{figure}

The unresolved central point-spread function shows spectral TAX phenomenology, i.e., two unrelated
thermal components subject to different absorption. The hard (1.5-7.3 keV) light curves (Fig.~\ref{fig4}) and 
spectra  (Fig.~\ref{fig5}) show frequent flares, while the soft (0.5-1 keV) component is steady. However, the 
soft emission gradually increased from 2004 to 2010, corresponding to an increase of the X-ray 
luminosity from 
$1.1\times 10^{29}$~erg~s$^{-1}$ (2004), to 
$1.2\times 10^{29}$~erg~s$^{-1}$ (2006), to
$1.8\times 10^{29}$~erg~s$^{-1}$ (2010), while the temperature and absorption did not change.

\begin{figure}
\centerline{
\includegraphics[angle=-90,width=11.cm]{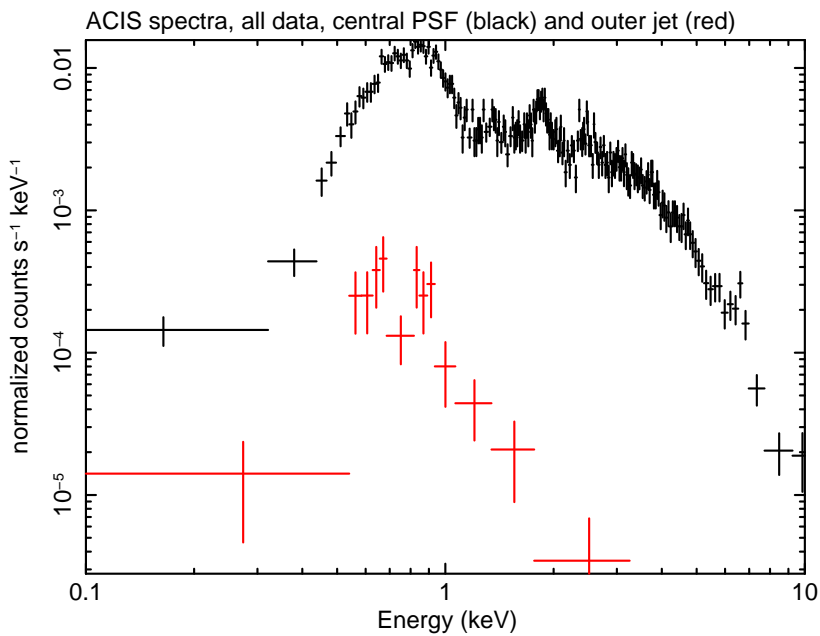}
}
\caption{Spectrum of central Chandra PSF (black) and the outer, resolved jet (red). All available Chandra data
were combined. Note similarity between jet spectrum and the soft peak of the central-PSF spectrum. }
\label{fig6} 
\end{figure}

The soft component indicates unusually low temperatures for X-ray sources in classical T Tauri stars. A 1-T thermal fit 
(using XSPEC/vapec) finds $T=3.7\pm 0.2$~MK (90\% errors), with 
$N_{\rm H}= (1.5\pm 0.4)\times 10^{21}$~cm$^{-2}$; the latter is lower by a factor of 2--4 than expected from 
visual extinction (1.4--3.3 mag, G07); but, both values agree with the spectral energy distribution 
of the resolved jet (Fig.~\ref{fig6}; $T \approx 2.7[2.0-3.8]$~MK), suggesting that the soft component 
originates  in the unresolved inner jet (G05). The soft image in Fig.~\ref{fig1} therefore shows 
only the jet - without the star!

\begin{figure}
\centerline{\includegraphics[angle=0,width=13.cm]{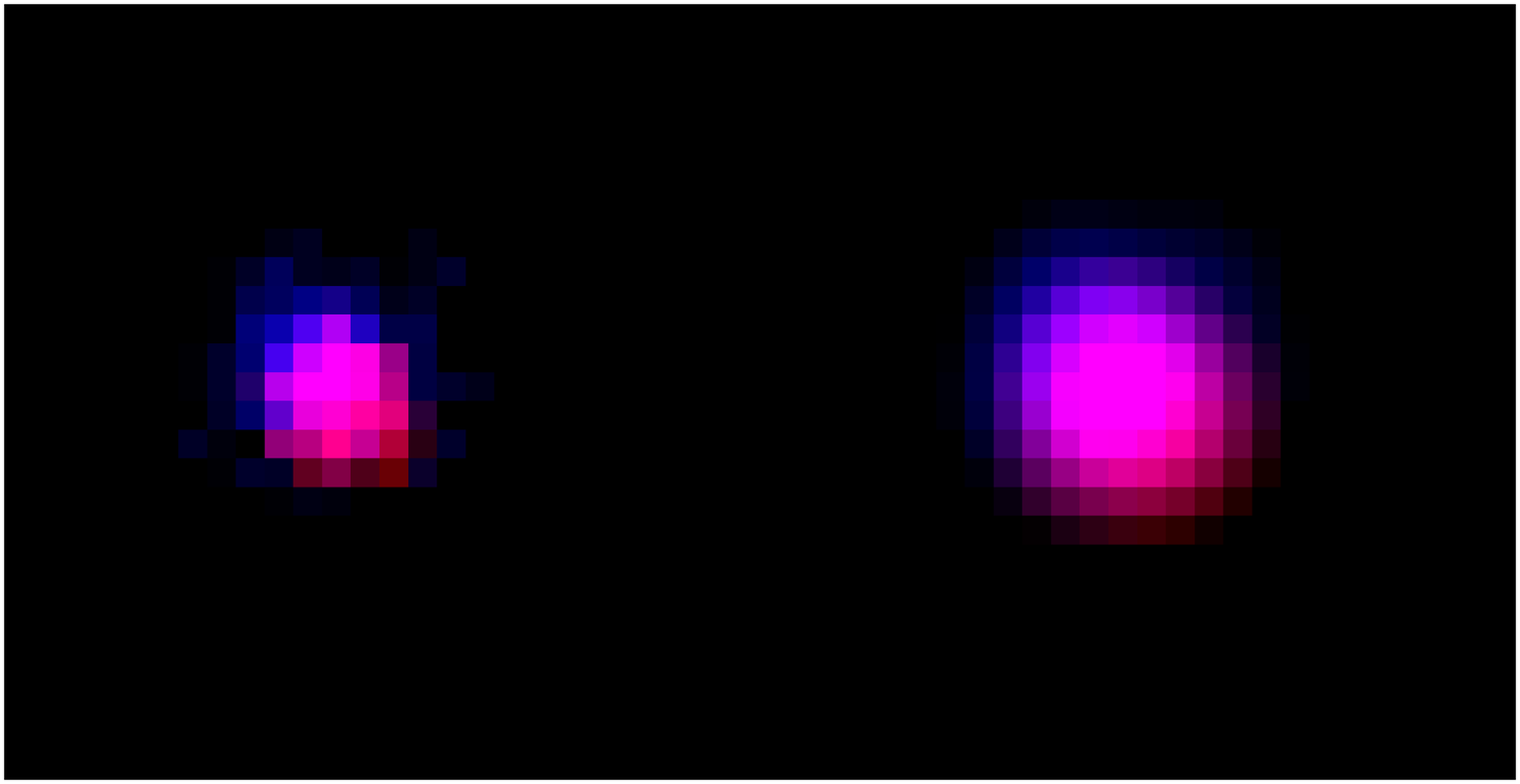}}
\caption{Two-color images illustrating the offset between the hard (1.5--7.3~keV, blue) and soft (0.45--1.1~keV, red) central PSFs for
          the combined 2010 observations. 
         {\it Left:} unsmoothed, 1 pixel = 0.125$^{\prime\prime}$. -- {\it Right:} smoothed using Gaussian of width = 3 pixels,
	         1 pixel = 0.125$^{\prime\prime}$.}
\label{fig7} 
\end{figure}

Further support comes from a systematic offset between the soft (jet?) and the 
hard (star) PSF centroids, already described by \citet{schneider08}. We used CIAO tasks dmstat 
(confined to the central PSF) and wavdetect on the SER-treated images to derive offsets of 
0.12$^{\prime\prime}$--0.16$^{\prime\prime}$ (dmstat), or 0.14$^{\prime\prime}$--0.20$^{\prime\prime}$ (wavdetect) in the jet direction -- 
offsets much smaller than Chandra's PSF. The 2-color image in Fig.~\ref{fig7} shows the hard 
(blue) vs soft (red) offset graphically, for the unsmoothed/smoothed PSF.

\section{Summary}
The DG Tau jet shows a rich X-ray phenomenology, including moving knots several arcseconds from the
star itself but apparently also jet emission regions much closer to the star and unresolved in 
Chandra images. Observations taken over 6 years show an apparent fading of the outer jet knots 
(and the near disappearance of the counter-jet), compatible with cooling models. On the other hand, 
the soft jet sources closer to the star has continuously brightened. We also measure X-ray jet motion 
in agreement with optical measurements. A more detailed interpretation of these data will be
given in a forthcoming publication.

\acknowledgements M.~A. acknowledges support from a Swiss National Science 
Foundation Professorship (PP002--110504).

\bibliography{guedel_m}

\end{document}